\documentstyle[11pt,aasms4]{article}

\newcommand\nion[2]{#1\,\lowercase{{\sc #2}}}
\newcommand\wave[1]{\mbox{$\lambda$#1\,\AA}}
\newcommand\iso[2]{${\rm ^{#2}}$#1}
\newcommand\eps[1]{log~$\varepsilon$(#1)}
\def\kmsec{\mbox{km~s$^{\rm -1}$}}
\def\teff{\mbox{T$_{\rm eff}$}}
\def\vt{\mbox{v$_{\rm t}$}}
\def\BmV0{\mbox{(B-V)$^{\rm o}$}}
\def\VmK0{\mbox{(V-K)$^{\rm o}$}}
\def\MV0{\mbox{M$_{\rm V}^{\rm o}$}}
\def\MV{\mbox{M$_{\rm V}$}}

\def\carbiso{\mbox{${\rm ^{12}C/^{13}C}$}}
\def\etal{\mbox{{\it et al.}}}
\def\eg{\mbox{{\it e.g.}}}
\def\ie{\mbox{{\it i.e.}}}

\def\third{{\it 3$^{rd}$}}
\def\oneonefive{{\rm HD~115444}}
\def\onetwotwo{{\rm HD~122563}}
\def\cs22892{{\rm CS~22892-052}}

\begin{document}

\title{The r-Process-Enriched Low Metallicity Giant HD~115444}

\author{Jenny Westin\altaffilmark{1,2},
Christopher Sneden\altaffilmark{1}, \\
Bengt Gustafsson\altaffilmark{2}, and
John J. Cowan\altaffilmark{3}
}

\altaffiltext{1}{Department of Astronomy and McDonald Observatory, 
University of Texas, Austin, TX 78712}
\authoremail{jenny@bellini.as.utexas.edu, chris@verdi.as.utexas.edu}

\altaffiltext{2}{Uppsala Astronomical Observatory, Box 515, 
S-751 20 Uppsala, Sweden}
\authoremail{be@astro.uu.se, bg@astro.uu.se}

\altaffiltext{3}{Department of Physics and Astronomy,
University of Oklahoma, Norman, OK 73019}
\authoremail{cowan@mail.nhn.ou.edu}

\vskip .5truein
\begin{center}
To appear in {\it The Astrophysical Journal}
\end{center}

\begin{abstract}

New high resolution, very high signal-to-noise spectra of ultra-metal-poor 
(UMP) giant stars \oneonefive\ and \onetwotwo\ have been gathered 
with the High-Resolution Echelle Spectrometer of the McDonald Observatory 
2.7m Telescope.
With these spectra, line identification and model atmosphere analyses
have been conducted, emphasizing the neutron-capture elements. 
Twenty elements with Z~$>$~30 have been identified in the 
spectrum of \oneonefive. 
This star is known to have overabundances of the neutron-capture 
elements, but it has lacked a detailed analysis necessary to compare with
nucleosynthesis predictions.
The new study features a line-by-line differential abundance comparison
of \oneonefive\ with the bright, well-studied halo giant \onetwotwo.
For \oneonefive, the overall metallicity is [Fe/H]~$\simeq$~--3.0.
The abundances of the light and iron-peak elements generally show the same
pattern as other UMP stars (e.g. overdeficiencies
of manganese and chromium, overabundances of cobalt), but the differential 
analysis indicates several nucleosynthesis signatures that are unique 
to each star.

Synthetic spectrum analyses reveal substantial overabundances of the 
heavier neutron-capture elements (Z~$\geq$~56; elements barium and beyond)
in \oneonefive.
Thus with [Eu/Fe]~$\simeq$~+0.9 for example, \oneonefive\ is a moderate 
version of the extremely neutron-capture-rich UMP giant 
CS~22892-052 ([Fe/H]~$\simeq$ --3.1, [Eu/Fe]~$\simeq$ +1.7).  
The abundance pattern of the heavier neutron-capture elements 
is consistent with scaled solar system $r$-process-only abundances 
(with little contribution from the s-process).
In \oneonefive, [Ba/Eu]~= --0.73, while in CS~22892-052, this ratio is --0.79.
Thus \oneonefive\ becomes the second UMP $r$-process-rich halo giant 
unambiguously identified from a very detailed abundance analysis.
Abundances of the lighter neutron-capture elements strontium, yttrium, and
zirconium are however nearly identical in \oneonefive\ and \onetwotwo.

Along with the heavier neutron-capture elements, the \wave{4019} line of 
\nion{Th}{ii} has been detected in \oneonefive,
yielding \eps{Th}~= --2.23~$\pm$~0.07. 
Comparing the observed thorium abundance in \oneonefive\, along with 
CS 22892--052, with other theoretical estimates of the time-zero 
abundance suggests an age for both of these UMP stars of 
15.6~$\pm$~4 Gyr, consistent with previous radioactive age estimates for 
CS 22892--052 and other Galactic and cosmological age determinations.

\end{abstract}


\section{Introduction}

The secrets of early Galactic nucleosynthesis are slowly being revealed from 
the chemical compositions of metal-poor stars.
Advances in analytical techniques have been applied to increasing
samples of fainter (thus more distant) stars, yielding more sharply defined
abundance trends with metallicity for several major element groups.
The general abundance patterns (and their exceptions) in stars of
the metallicity range 0.0~$\geq$~[Fe/H]~$\geq$~--2.5\footnote
{We adopt the usual spectroscopic notations that
[A/B]~$\equiv$~log$_{\rm 10}$(N$_{\rm A}$/N$_{\rm B}$)$_{\rm star}$~--~log$_{\rm 10}$(N$_{\rm A}$/N$_{\rm B}$)$_{\odot}$, and
that \eps{A}~$\equiv$~log$_{\rm 10}$(N$_{\rm A}$/N$_{\rm H}$)~+~12.0,
for elements A and B. Also, metallicity will be assumed here to be
equivalent to the stellar [Fe/H] value.}
are now reasonably well established for most spectroscopically
accessible elements with atomic numbers Z~$\leq$~30.
A good recent summary of these trends has been given by Cayrel 
(1996\markcite{Ca96}).

Elements with Z~$>$~30, comprising more than half of the Periodic
Table, are synthesized almost exclusively in neutron-capture fusion
reactions.
Abundance trends with metallicity are on firm ground for
relatively few of these neutron-capture elements, primarily
due to their small abundances and lack of strong transitions available 
to ground-based spectroscopy for most of these elements.
Nevertheless some general statements may be made from the extant observations.  
First, the ``bulk'' levels of neutron-capture elements 
vary greatly with respect to the Fe-peak elements in the lowest
metallicity stars ([Fe/H]~$\lesssim$~--2.5), hereafter called
ultra-metal-poor (UMP) stars.
The neutron-capture-to-Fe ratios can scatter by more than a factor
of 100 from star to star at a given metallicity 
(Gilroy \etal\ 1988\markcite{GSPC88}, McWilliam \etal\ 
1995\markcite{MPSS95}, Burris \etal\ 1999\markcite{Betal99}).
Second, the Ba/Eu abundance ratio slowly declines with decreasing 
metallicity, reaching [Ba/Eu]~$\simeq$~--0.8 at [Fe/H]~$\simeq$~--3 
(McWilliam 1998\markcite{Mc98}, Burris \etal\ 1999\markcite{Betal99}).
Europium is synthesized most easily in conditions of extremely
high neutron fluxes (during the so-called {\it r[apid]}-process)
while barium is most easily made in very low neutron-flux environments
(during the {\it s[low]}-process).
The decline in [Ba/Eu] is thus taken as evidence for the dominance of
a Type~II supernova-induced $r$-process in producing neutron-capture 
elements early in the Galaxy's nucleosynthesis history.
Third, the abundances of ``lighter'' Sr--Y--Zr neutron-capture elements 
(Z~=~38--40) do not simply scale with those of the ``heavier'' elements
Ba--... (Z~$\geq$~56).
Nor do these lighter neutron-capture elements fit easily into a single
nucleosynthesis scheme (\eg, Cowan \etal\ 1995\markcite{CBSPM95}).

Confidence in the $r$-process origin of the heavier neutron-capture 
elements in UMP stars does not rest solely with the many 
determinations of [Ba/Eu] ratios, which are still subject to 
reasonably large observational uncertainties.
Instead, the most convincing evidence has come from the detailed
abundance distributions of neutron-capture elements in those 
UMP stars that are bright enough to permit very high resolution,
high signal-to-noise spectroscopy (so that very weak neutron-capture
element transitions may be detected), or those that have anomalously 
large [n-capture/Fe] ratios (so that the neutron-capture transitions
are detected in high contrast with features of other elements). 
Such detailed abundance distributions (at least 10 neutron-capture 
elements) exist for only a few UMP stars. 
For example, Sneden \& Parthasarathy (1983\markcite{SP83}) published a 
detailed neutron-capture element study of the bright UMP giant \onetwotwo. 
These elements are actually underabundant with respect to Fe in
\onetwotwo\ (\eg, [Eu/Fe]~$\simeq$~--0.4).
Another UMP giant star, \cs22892, originally 
identified in the objective prism survey of the halo by Beers, 
Preston, \& Shectman (1985, 1992)\markcite{BPS85}\markcite{BPS92}, 
is a much fainter star at V~=~13.1.
But it has been the target of several recent abundance studies
(Sneden \etal\ 1994\markcite{SPMS94}, 1996\markcite{SMPCBA96};
Norris, Ryan, \& Beers 1997a\markcite{NRB97a}) because it has the largest 
overabundances of neutron-capture elements (\eg, [Eu/Fe]~$\simeq$~+1.7) 
of any UMP star outside of the CH-star domain.
The common link between the neutron-capture abundance mixes of \onetwotwo, 
\cs22892, and many other UMP stars studied by Gilroy \etal\ (1988) and 
McWilliam \etal\ (1995) is the $r$-process signature among the heavy elements.
The relative abundance patterns are remarkably similar, and so far
are indistinguishable from the abundance pattern of
the $r$-process-only parts of solar-system material (K{\"a}ppeler, Beer,
\& Wisshak 1989\markcite{KBW89}, Sneden \etal\ 1996\markcite{SMPCBA96}).

An important aspect of neutron-capture element studies in
UMP stars is nucleocosmochronometry. 
This is the determination of the ages of halo stars (commonly assumed 
to be nearly equal to the galactic age) from abundances of elements
that radioactively decay with long half-lives.
The technique was pioneered by Butcher (1987)\markcite{Bu87} and has 
been applied in practice mainly to the spectroscopically accessible 
element thorium. 
This element is synthesized exclusively in the $r$-process
and decays with $\tau_{\onehalf}$~=~14.0~Gyr.
But the single strong \nion{Th}{ii} feature routinely available to
ground-based spectroscopy is part of a complex atomic and molecular
transition blend at \wave{4019}, as illustrated in spectra published by
Morell, K{\"a}llander, \& Butcher (1992)\markcite{MKB92}, 
and Fran\c{c}ois, Spite, \& Spite (1993)\markcite{FSS93}.
Thorium abundances accurate enough to be useful for cosmochronometry 
can only be obtained when [Th/Fe]~$\gg$~0, so that the \nion{Th}{ii}
line dominates the other \wave{4019} blend components.
Stars with high thorium abundances will also have large abundances
of other neutron-capture elements.
That is fortunate, because determinations of ages from just 
[Th/Eu] or [Th/Nd] ratios have significant uncertainties due to the
big atomic number gap between thorium and the other elements.
A thorium decay age is computed from the difference between 
the presently observed [Th/n-capture] ratio and the estimated value 
this ratio had immediately after supernova synthesis. 
Thus in order to solidify observational aspects of thorium 
cosmochronometry, many abundances throughout the Z~=~60--90 
range must be determined in many stars. 
This means that UMP stars with neutron-capture element overabundances 
must be sought out and subjected to extensive analyses.

The star \oneonefive\ was first noted in Bond's (1980)\markcite{Bp80} objective 
prism search for low metallicity halo giants.
But its unusual spectrum was discovered by Griffin \etal\
(1982\markcite{GGGV82}, hereafter GGGV82), who obtained moderately
high resolution photographic spectra
of \oneonefive\ after it failed to yield a sensible measurement
with Griffin's (1967)\markcite{Gr67} photoelectric radial velocity
spectrometer.  
Their analysis of \oneonefive, performed differentially with respect 
to \onetwotwo, concluded that \oneonefive\ probably was more 
metal-poor than \onetwotwo\ ([Fe/H]~$\simeq$~--3.0 versus --2.7). 
They also found all six neutron-capture elements detected in both stars 
to be more abundant in \oneonefive\ than in \onetwotwo. 
However, the heaviest neutron-capture elements Ba, La, and Eu were 
far more overabundant than the lighter elements Sr, Y, and Zr.
GGGV82 labeled \oneonefive\ as ``a barium star of extreme Population~II'',
but noted that the $r$-process element Eu seemed nearly
as overabundant as the $s$-process element Ba.
Gilroy \etal\ (1988)\markcite{GSPC88} included \oneonefive\ in their 
neutron-capture abundance survey of Bond (1980)\markcite{Bo80} giants.
Their results, determined from an echelle/CCD spectrum, 
did not indicate an enhancement of barium.
Instead, they derived a large europium abundance leading to 
[Ba/Eu]~$\simeq$~--1.0, indicative of an $r$-process origin for 
the heavier neutron-capture elements.

Recently, Sneden \etal\ (1998a)\markcite{SCBT98} obtained HST GHRS
data in three UV spectral regions of \oneonefive, and they detected (for the 
first time in any UMP star) some transitions of the very heavy 
neutron-capture elements osmium, platinum and lead.
Unfortunately, their interpretation of the abundances of these elements 
was somewhat compromised by the uncertain (possibly conflicting) abundances 
of other neutron-capture elements from the GGGV82 and Gilroy 
\etal\ (1988)\markcite{GSPC88} studies.
In this paper our goals are: {\it a)} to clear up questions on the status 
of \oneonefive\ remaining from the earlier ground-based studies; 
{\it b)} to expand the number of detected neutron-capture elements 
in this unusual star; {\it c)} to connect in a consistent way 
its ground-based and HST-based abundances; and {\it d)} to make an age 
estimate for the \oneonefive\ material from its thorium abundance.
We present a comprehensive abundance study of \oneonefive\
from newly acquired high resolution, high signal-to-noise spectra
spanning the near-UV to red spectral regions.
Following the lead of GGGV82, we have analyzed \oneonefive\ with respect to
\onetwotwo, because the atmospheric parameters of these two halo UMP red
giants are so similar.
In succeeding sections we describe the data acquisition and reduction,
discuss the differential model atmosphere calculations and abundance 
derivation of these stars, and then concentrate on implications of 
the neutron-capture element abundances in \oneonefive.

\section{Observations, Reductions, and Equivalent Widths}

In 1997 we obtained high resolution, high signal-to-noise spectra
of \oneonefive\ and \onetwotwo\ with the McDonald Observatory 2.7m H.~J. Smith
telescope and the ``2d-coud\'e'' cross-dispersed echelle spectrograph
(Tull \etal\ 1995)\markcite{TMSL95}.
Basic data for these two stars are given in Table~\ref{tab1-basic}.
We also obtained companion spectra of a tungsten lamp to use in
flat-fielding the raw data frames, a Th-Ar lamp for eventual wavelength
calibration of the extracted spectra, and a rapidly rotating hot star
for telluric absorption line cancelation.
The spectrograph was configured such that with an entrance aperture width
of 1.2~arcsec in the dispersion dimension, the 2-pixel
spectral resolving power at the Tektronix 2048$\times$2048 CCD
was R~$\equiv$~$\Delta\lambda/\lambda$~$\simeq$ 60,000. 
The spectra spanned the wavelength range
3600~\AA~$\leq$~$\lambda$~$\leq$~10000~\AA. 
Blueward of 5900~\AA\ the
spectral coverage was complete, but redward of this wavelength the
free spectral range of the echelle orders was too large to be contained
on the CCD, thus producing wavelength gaps in the recorded spectra.
The S/N ratios of the spectra were of course wavelength-dependent,
such that at \wave{6000}, S/N~$\simeq$~500; at \wave{5000}, 
S/N~$\simeq$~400; and at \wave{4000}, S/N~$\simeq$~200.

The raw data frames were processed in standard ways using IRAF
echelle reduction routines.
After extraction of wavelength calibrated echelle orders, cancelation of
telluric spectral features in some of the orders was accomplished with 
specialized software (Fitzpatrick \& Sneden 1987\markcite{FS87}).
Sample spectra of \oneonefive\ and \onetwotwo\ are shown in 
Figure~\ref{fig1-sample}.
In general, absorption features in \onetwotwo\ are somewhat stronger
than in \oneonefive. This makes sense, given previous claims  
that \onetwotwo\ is more metal-rich and has a lower
effective temperature than \oneonefive. 
However, as seen in the upper panel, the transitions of the ``heavy'' 
(Z~$\geq$~56) neutron-capture elements  La and Eu  
are clearly much stronger in \oneonefive. 
On the other hand, the transitions displayed in the lower panel 
shows no corresponding enhancements of the ``lighter'' neutron-capture 
elements Zr and Sr in \oneonefive.
  
Equivalent widths (EWs) were measured in order to derive model atmosphere
parameters and to determine abundances for those elements possessing
many unblended lines in our spectra.
We measured the EWs with the IRAF task {\it splot}, using a Gaussian 
assumption for the line profiles. 
Two continuum points were marked and a single best-fit line profile 
was displayed and examined for obvious blends. 
Listed in Table~\ref{tab2-EWs} are the measured EWs of \oneonefive\ 
and \onetwotwo\ together with the excitation potentials (EPs) and 
transition probabilities (log gf's) of the lines.
For most lines we adopted the log gf's employed by McWilliam \etal\ (1995)
and Sneden \etal\ (1996); see those papers for the sources of these data.
We also supplemented some log gf's from Fuhr \& Wiese (1996\markcite{fur96}).
Laboratory gf values have been employed throughout this study; none
have been derived from a solar spectral analysis.
For those lines that were analyzed using synthetic spectra,
the entry {\it syn\/} is used in Table~\ref{tab2-EWs} instead of an EW value.

We compared our measured EWs with those of previous studies of 
\oneonefive\ and \onetwotwo\ in the literature. 
Taking differences in the sense {\it this work minus others}, 
the mean differences with the measurements by GGGV82 are for \oneonefive,
$<\delta$EW$>$~= --15.2~$\pm$~1.8~m\AA\ ($\sigma$~=~15.9~m{\rm \AA}, 
79 lines in common); and for \onetwotwo, 
$<\delta$EW$>$~= --15.9~$\pm$~1.8~m\AA\ ($\sigma$~=~16.2~m\AA, 81 lines). 
The large line-to-line scatter and the overall offset are undoubtedly
due to the use of photographic plates in the 1982 study. 
Those data had significantly lower S/N and resolution than do 
the present spectra.
On the other hand, Sneden \& Parthasarathy's (1983) EWs of \onetwotwo\
agree well with our measurements: 
$<\delta$EW$>$~= +0.10~$\pm$~0.6~m\AA, ($\sigma$~=~6.2~m\AA, 108 lines). 
Gilroy \etal\ (1988) published data for \oneonefive\ and \onetwotwo\ 
as part of a larger survey of stars, and the compared EWs are in 
reasonable accord: for \oneonefive\ 
$<\delta$EW$>$~= --4.2~$\pm$~2.8~m\AA, ($\sigma$~=~10.9~m\AA, 15 lines), 
and for \onetwotwo, 
$<\delta$EW$>$~= --2.5~$\pm$~4.4~m\AA, ($\sigma$~=~12.5~m\AA, 8 lines). 
The standard deviations are dominated by a couple of very discrepant lines. 
Gratton \& Sneden (1990)\markcite{GS90} also have a small number 
of lines in common with our study, but only for \onetwotwo. 
Their values agree well with our measurements:
$<\delta$EW$>$~= +1.3~$\pm$~1.4~m\AA, ($\sigma$~=~4.65~m\AA, 11 lines). 
Since our new spectra are of higher S/N and resolution than (most of) the 
compared studies, the encountered differences were not unexpected. 
On the whole our EWs agree reasonably well with those in the literature.

\section{Model Atmospheres and ``Ordinary'' Abundances}

\subsection{Derivation of Fundamental Model Parameters}

We first derived stellar model atmospheres and abundances for \oneonefive\
and \onetwotwo\ in two independent analyses. 
We then followed the spirit of the GGGV82 study by using 
\onetwotwo\ as a template star to derive more accurate {\it differential\/} 
model atmosphere parameters for \oneonefive.
Results of the independent analyses of each star will be labeled 
``absolute'', while the term ``relative'' will be reserved for results
of the differential analysis.

Trial stellar atmosphere models with different values of input parameters 
(\teff, [M/H], log~g, and \vt) were generated by a new version 
of the model atmosphere code MARCS.
These atmospheres are one-dimensional, plane-parallel, flux-constant 
LTE models computed using the opacity sampling scheme with 
21,000 wavelength points to approximate the relevant gas opacity properties. 
MARCS is a modern version of the code described by Gustafsson \etal\ 
(1975)\markcite{GBEN75} and Edvardsson \etal\ (1993\markcite{EAGLNT93}), 
and it is subject to ongoing refinements. 
The code, together with the sources of its opacity data, 
will be described elsewhere.
For the current MARCS implementation, modern data for metallic and 
molecular lines and continua were compiled from many different sources.
The mixing-length approximation was used for convection, with parameters
$\alpha=\ell/H_{\rm p}=1.5$, $\nu=8$ and $y=0.076$ (Henyey \etal\ 
1965)\markcite{HVB65} for our models. 
An enhancement of 0.4 dex of the $\alpha$-process elements (from
oxygen through calcium) was also assumed. 
 
Final choice among possible trial model atmospheres was made 
by using them and the measured EWs as input for the current version 
of the LTE line analysis program MOOG (Sneden 1973)\markcite{Sn73}.
The model parameters were determined as follows.
{\it Effective temperature\/}: \teff\ was iteratively adjusted until the 
trend of \nion{Fe}{i} abundance with excitation potential was minimized. 
{\it Microturbulence\/}: \vt\ mostly affects the abundances derived 
from strong (saturated) lines. 
Therefore a plot of abundance vs. EW will show a slope if the microturbulence 
is in error; \vt\ was iteratively determined by minimizing such slopes.
{\it Gravity\/}: To find the appropriate log~g we required 
that the abundance determined from ionized species be equal 
to the abundance derived from neutral species (Fe and Ti were used).
{\it Metallicity\/}: the model metallicity was constrained to
be close to the derived [Fe/H] value from the Fe line analysis.
The model atmosphere parameters for \oneonefive\ and \onetwotwo\ derived from 
these criteria are listed in Table~\ref{tab1-basic}.
The model metallicities, [M/H] = --2.90 for \oneonefive\ and --2.70 for
\onetwotwo, are both within 0.1~dex of our final [Fe/H] values
(--2.99 and --2.74, respectively).
 
To illustrate the \teff\ and \vt\ derivations, Figure~\ref{fig2-felines} 
shows abundances of \nion{Fe}{i} lines derived with the final model 
of \oneonefive\ plotted against excitation potential 
(EP) and the logarithm of the reduced width (log~EW/$\lambda$).
Trends of abundances with EP and log~EW/$\lambda$ have been minimized. 
The mean \nion{Fe}{i} abundances agree to the line-to-line scatter limit 
($\sigma \simeq $ 0.1 dex) for lines ranging over nearly 5 eV in EP and 
over two orders of magnitude in EW.

In order to investigate internal abundance uncertainties we computed the
dependence of derived abundances for \oneonefive\ on each of the 
stellar atmosphere parameters.
Relative to the final model, [M/H] was increased by 0.5, 
\teff\ by 150 K, log~g by 0.3, and \vt\ by 0.3 \kmsec. 
The abundance changes with these parameter excursions are given
in Table~\ref{tab3-errors}.
Some trends are clear, such as the inevitable rise in all abundances
as \teff\ increases.
A metallicity change hardly alters the abundances. 
Gravity changes affect neutral and ionized species abundances
in opposite directions.
Larger microturbulence values will decrease the abundances, but this 
parameter affects just the strong (saturated) lines, and thus is not a 
large effect for most species in these generally weak-lined stars.
Exceptions to this statement are those species exhibiting only saturated
lines (\eg, \nion{Sr}{ii}, \nion{Ba}{ii}).
Even here, the entries of Table~\ref{tab3-errors} constitute 
a worst-case set of internal uncertainties.
Many {\it abundance ratios} ({\it i.e.}, [X/Y] for elements X and Y), 
will be substantially smaller, especially when comparing abundances
from the same ionization states.
Furthermore, derivation of \teff\ and log~g through excitation/ionization
equilibria considerations inevitably couples them: higher values of
\teff\ also force higher values of log~g, which partially cancels any
abundance changes.

The main sources of line-to-line abundance scatter within a species for each 
star are line blending, log~gf uncertainties, and EW measurement errors.
These errors can be substantially reduced by computing relative
abundances between the two stars.
We show this in Figure~\ref{fig3-deltalines} by plotting line-by-line 
abundances differences $\Delta$~\eps{X}~$\equiv$ 
\eps{X}$_{\rm \oneonefive}$~--~\eps{X}$_{\rm \onetwotwo}$
with respect to EP and EW,
using the abundances generated with the final models for each star. 
Not only are the slopes negligible (implying well-determined \teff\ 
and \vt\ differences), but the line-to-line scatters have shrunk from 
$\simeq$~0.1 to $\simeq$~0.05~dex. 
In these relative abundances the uncertainties in 
individual log~gf values have been explicitly eliminated. 
Additionally, errors due to unaccounted-for line blending should 
affect the derived abundances in the stars in approximately the same 
way if the blending agents are transitions of elements with Z~$\leq$~30.
Thus we expect a further reduction of error in the relative quantities. 
Some line measurement errors may also be reduced, to the extent that 
continua have been set and line profiles measured in consistent ways
for the two stars.
Notice in the lower left-hand panel of Figure~\ref{fig3-deltalines}
that the line-to-line scatter in $\Delta$~\eps{Fe} is largest 
for the stronger lines (log EW/$\lambda > -4.8$). 
These lines lie on the flat and damping portions of the curve-of-growth,
where small EW errors can lead to large abundance errors. 
But even for these lines the relative Fe abundance is more accurately
determined than the absolute abundance of each star.

Choice of model atmosphere code affects the derived stellar parameters
and absolute abundance scales.
We tested the magnitude of this uncertainty source by deriving atmosphere 
parameters for the two stars with models interpolated within a recent grid of 
atmospheres (Kurucz 1992)\markcite{Ku92} generated with the ATLAS code.
The resulting models are approximately 125~K hotter in \teff, about
0.2~dex higher in log~g, but require no obvious changes in microturbulent 
velocities \vt.
The derived abundances average $\simeq$0.06~dex greater with these models.
However, model parameter and individual abundance {\it differences} between 
the two stars are virtually unchanged by this switch in model codes.

\subsection{External Comparisons}

Our derived model parameters agree well with previous estimates, 
especially in the differential sense.
There are many more published analyses of \onetwotwo\ in the
literature than of \oneonefive.
The most recent ``Catalogue of [Fe/H] Determinations'' (Cayrel de Strobel
\etal\ 1997)\markcite{CSFRF97} lists 24 spectroscopic analyses of
\onetwotwo.
Taking only those analyses based on full model atmosphere calculations,
and attempting to eliminate those that adopt most model parameters
from other papers, the remaining literature studies on average
report \teff~$\simeq$~4600~K, log~g~$\simeq$~1.1, and [Fe/H]~$\simeq$~--2.65.
The corresponding values from the present study are 4500~K, 1.3, and --2.74.
Our derivation of a slightly lower metallicity for \onetwotwo\
probably is a consequence of our lower \teff\ estimate.

It is more relevant here to discuss those literature studies that treat
both of our stars, beginning with effective temperature estimates.
Taking differences between the two stars in the sense {\it \oneonefive\ 
minus \onetwotwo}, our temperature analyses yielded 
$\Delta$(\teff)~= 4650~--~4500~= +150~K. 
Three spectroscopic studies have dealt with both of these stars.
GGGV82 found $\Delta$(\teff)~= 4800~--~4600~= +200~K (and also pointed 
out that the difference in effective temperatures is more certain 
than the individual values). 
Gilroy \etal\ (1988) concurred with these values.
Gratton \& Ortolani (1984) derived $\Delta$(\teff)~= 4850~--~4680~= +170~K. 
Recently, major efforts have been devoted to calibrating broad-band 
photometric indices with \teff.
The V--K color is nearly a pure function of temperature,
and using Di Benedetto's (1998)\markcite{DB98} second-order polynomial 
fit of \teff\ to V--K and the colors of Table~\ref{tab1-basic}, 
$\Delta$(\teff$_{V-K}$)~= 4707~--~4575~= +132~K.
The B--V color is a more complex function of both \teff\ and [Fe/H],
due to the large B-band line blanketing in cool stars.
Alonso \etal\ (1996)\markcite{AAM96} derived a relation for \teff\ in 
terms of B--V and [Fe/H] variables for main sequence stars,
and Alonso (1999)\markcite{Al99} has provided a similar relation for
giant stars.
This new formula and the data of Table~\ref{tab1-basic}
yield $\Delta$(\teff$_{B-V}$)~= 4853~--~4717~= +136~K.
None of the spectroscopic or photometric estimates of \teff\ differences
would claim better than $\simeq$50~K accuracy; thus our newly derived
$\Delta$(\teff)~= +150~K between the two stars agrees well 
with the other values.
The absolute \teff\ scale is less certain, and we note that our temperatures
are typically $\sim$100~K lower than others.
This small discrepancy may partly be related to our use of MARCS model
atmospheres (\S 3.1). 
But it does not affect any of the basic conclusions 
of this paper and will not be considered further here.

The derived gravity difference is 
$\Delta$(log~g)~= 1.50~--~1.30~= +0.20. 
Similar results were obtained by GGGV82, 
$\Delta$(log~g)~= 1.6~--~1.2~= +0.4, 
and by Gratton \& Ortolani (1984), 
$\Delta$(log~g)~= 1.60~--~1.50~= +0.10. 
The gravity difference derived by Gilroy \etal\ (1988) is somewhat larger,
$\Delta$(log~g)~= 2.0~--~1.2~= +0.8, but the agreement among the 
other studies indicates the Gilroy \etal\ log~g value for
\oneonefive\ is probably too large.

Our estimation of the microturbulence difference is 
$\Delta$(\vt)~= 2.1~--~2.5~= --0.4~\kmsec. 
Previous results for the microturbulence vary. 
GGGV82 found the same value in both stars, \vt~=~2.6~\kmsec, 
as did Gratton \& Ortolani (1984), \vt~=~3.0~\kmsec. 
Gilroy \etal\ (1988) derived 
$\Delta$(\vt)~= 2.2~--~2.3~= --0.1~\kmsec.
These comparisons are not very enlightening, but do suggest that the
microturbulence values for the two stars are not very different.

We determined a relative metallicity $\Delta$~\eps{Fe}~= 
--2.99~--~--2.74~= --0.25 (and --0.24 from the line-by-line differential 
analysis), taking the mean of neutral and ionized species 
(Table~\ref{tab4-abund}).  
This difference is not easily anticipated by past spectroscopic studies.
Previously estimated [Fe/H] metallicities for \oneonefive\ and \onetwotwo\ 
are: --2.95 and --2.75, respectively (GGGV82); --2.0 and --2.0 
(Gratton \& Ortolani 1984); --2.5 and --2.6 (Gilroy \etal\ 1988).
These literature metallicity values bear little relationship to the
microturbulent velocities determined in each study, and the
analysis techniques are dissimilar enough to not warrant further
exposition here.
Some calibrated photometric indices provide further estimates of
the metallicities:  [Fe/H]~= --2.7 for \oneonefive\ and --2.6 for 
\onetwotwo\ (Bond 1980, $uvby$); --2.56 and --2.47 (Anthony-Twarog
\& Twarog 1994\markcite{ATT94}, $uvby$); --2.83 and --2.58
(Clari\'a \etal\ 1994; DDO $\delta$4548).
Thus taken as a whole, the previous spectroscopic and photometric 
investigations do not disagree with the modest metallicity difference 
found in the present work.

We performed a further check on the spectroscopically-derived gravity,
by computing an ``evolutionary'' gravity using the standard formula
$$
\log{\rm g_{evol}} = -12.50 + 0.4\left({\rm M_V} + {\rm BC}\right) + 
\log \mathcal{M} + 4\log \teff\
$$
\noindent which combines Newton's gravitation law with the Stefan-Boltzmann 
law, eliminating the stellar radius~R.
We assumed masses equal to the standard mass of stars near the 
Population~{\sc II} turnoff, $\mathcal{M}$~=~0.85$\mathcal{M}$$_{\odot}$.
With the derived \teff's, we used the calibrations of Montegriffo \etal\ 
(1998)\markcite{MFOF98} to determine bolometric corrections BC~=~--0.50 
for \oneonefive\ and --0.59 for \onetwotwo.
There are a few photometric estimates of \MV\ for the two stars
in the literature.
Anthony-Twarog \& Twarog (1994) derived \MV~=~--0.49~$\pm$~0.4 for 
\oneonefive\ and \MV~=~--1.24~$\pm$~0.4 for \onetwotwo, while Hanson \etal\ 
(1998) suggest \MV~=~--0.8~$\pm$~0.5 and \MV~=~--1.2~$\pm$~0.5 (from
a renormalization of the \MV\ estimates of Bond 1980).
Using the Anthony-Twarog \& Twarog absolute magnitudes in the above 
equation yield log~g$_{evol}$~=~1.70~$\pm$~0.20 for \oneonefive\ and 
log~g$_{evol}$~=~1.31~$\pm$~0.20 for \onetwotwo, and the Hanson \etal\ 
magnitudes yield log~g$_{evol}$~=~1.6~$\pm$~0.2 and 
log~g$_{evol}$~=~1.3~$\pm$~0.2, respectively.
This is in good agreement with our log~g estimates.

Hipparcos parallaxes (ESA 1997)\markcite{ESA97} exist for both stars: 
3.55~$\pm$~1.12 milliarcsec for \oneonefive\ and 3.76~$\pm$~0.72 milliarcsec 
for \onetwotwo --- almost the same values. 
But the apparent V of \onetwotwo\ is 2.78 magnitudes smaller 
than that of \oneonefive. 
The standard distance modulus relation 
(${\rm M_V = V + 5\log p + 5 - A_V}$) then implies nearly the same difference 
in absolute magnitudes in the absence of significant ISM extinction.
This is surprising since these stars, both red giants with very 
similar derived atmosphere parameters, would be expected to differ 
little in luminosity (as they indeed do according to literature estimates).
Using the evolutionary gravity equation for \onetwotwo, and adopting 
Bond's (1980) estimate that A$_{\rm V}$~=~0.0, we calculated from the 
Hipparcos parallax that log~g$_{\rm Hipp}$~=~1.42~$\pm$~0.21, 
very close to our spectroscopic log~g~=~1.30. 
The uncertainty in log~g$_{\rm Hipp}$ was computed from the maximum
and minimum values permitted by the uncertainties in the quantities 
upon which it depends.
Repeating the same calculation for \oneonefive\ (for which Bond also
estimated A$_{\rm V}$~=~0.0) gave log~g$_{\rm Hipp}$~=~2.58~$\pm$~0.33;
this is far from our estimate (in agreement with others) 
of log~g~=~1.50.
From this discussion and the results of literature comparisons, 
probable statistical biases make it is very risky to use such small 
parallaxes with large relative errors (34\% for \oneonefive).

\subsection{Light and Fe-peak Elements}

With the final model atmospheres, we determined abundances for nearly all
elements with Z~$\leq$~30 from their measured EWs. 
The exception was manganese, for which we treated the strong \nion{Mn}{i} 
lines in synthetic spectrum calculations, obtaining the hyperfine
substructure components from Kurucz's (1999)\markcite{Ku99} web site.
Mean \eps{X} abundances and [X/Fe] relative abundance ratios are given
for both stars in Table~\ref{tab4-abund}.
For a species with at least three available transitions in a stellar spectrum,
the standard deviation of the sample $\sigma$ quoted in Table~\ref{tab4-abund} 
is simply the internal scatter of individual measurements.
The mean of those values is $<\sigma>$~= 0.10 for \oneonefive\ 
(24 measures), and 0.12 for \onetwotwo\ (15 measures).
We take the average of these numbers as representative of the scatter 
of individual values of a given abundance for our stars.
Thus for a species with only one transition we entered 0.11 as
$\sigma$ in Table~\ref{tab4-abund}, and for a species with two transitions 
we entered the greater of 0.11 and the measured $\sigma$.
In subsequent discussions we will use these $\sigma$~values for the
abundance uncertainties, rather than the more commonly employed
standard deviations of the means ($\sigma$/$\sqrt{\# lines}$).
Adoption of the more conservative $\sigma$ allows many remaining 
uncertainties of scale (from choices of model atmosphere grids, 
transition probabilities, etc.) to be effectively accounted for.
These $\sigma$'s are also far more realistic abundance uncertainty
estimates for species with small numbers of measured transitions.

In Figure~\ref{fig4-metallicity} we show the individual abundances [M/H] 
for all the elements.
The horizontal solid line at [M/H]~=~--2.99 in this figure represents 
the mean [Fe/H] for \oneonefive, and the horizontal dotted line
at [M/H]~=~--2.74 is the mean [Fe/H] for \onetwotwo.
For clarity in this figure the abundance $\sigma$'s from Table~\ref{tab4-abund}
are not plotted, but of course they should be remembered when comparing
the abundances of different elements.
The data of Table~\ref{tab4-abund} and Figure~\ref{fig4-metallicity} 
demonstrate that the light and Fe-peak abundances (elements with 
Z~$\leq$~30 shown in the upper panel) in \oneonefive\ and \onetwotwo\ 
generally follow the abundance patterns of UMP stars 
(\eg, McWilliam \etal\ 1995, Ryan, Norris, \& Beers 1996\markcite{RNB96}, 
Cayrel 1996). 
That is, O, Mg, Si and Ti are enhanced relative 
to iron, while Al, Cr, Mn, and Cu are all deficient. 
In fact, nearly all of these abundance ratios of our program stars are 
within 0.1~dex of the mean ratios summarized in Cayrel's Figure~4.

However, some noteworthy anomalies exist in the abundance patterns.
To see this more clearly, in Table~\ref{tab4-abund} we also give 
for all species the abundance differences between the two stars.
These differences, like those discussed in \S3.1, were computed
on a line-by-line basis.
As before we define $\Delta$~\eps{X}~$\equiv$ 
\eps{X}$_{\rm \oneonefive}$~--~\eps{X}$_{\rm \onetwotwo}$.
The single-measure standard deviation in the relative abundance 
$\sigma$($\Delta$~\eps{X}) is usually much smaller than those of
the absolute abundances $\sigma$(\eps{X}).
For species with three or more transitions,
$<\sigma$($\Delta$~\eps{X})$>$~=~0.06~$\pm$~0.01 from 13 values, 
excluding the anomalously large scatter in \nion{Ba}{ii}.
Therefore 0.06 is entered as $\sigma(\Delta$~\eps{X}) for a species with
only one transition, and the greater of 0.06 and the measured value
is entered for a species with two transitions.

The abundance differences from Table~\ref{tab4-abund} are displayed
in Figure~\ref{fig5-diffabund} (not every element was detected in
\onetwotwo).
For the elements with detected transitions of both neutral and 
ionized species (Ti, V, and Fe), we plot a single point representing 
the mean differences.
The vast majority of elements with Z~$\leq$~30 (13 out of 18) have 
$\Delta$~\eps{X} values within 2$\sigma$ of 
$\Delta$~\eps{Fe}~=~--0.24. 
The abundances of these elements thus simply scale with the metallicity
difference between the two stars.
The most notable exceptions are the large relative C, Ti excesses 
and the significant relative Mn underabundance in \oneonefive.
For elements with Z~$>$~30 our results, supporting and extending those 
of GGGV82 and Gilroy \etal\ (1982), show only a small enhancement of
the lighter neutron-capture elements in \oneonefive\ but about a factor of
ten difference among the heavier elements.
We will now comment on these and other individual abundances: 

{\it The CNO-group: carbon.} 
The CNO elements were not of primary interest in this study, so to 
determine carbon and nitrogen abundances we simply adopted the 
atomic and molecular linelists from the Kurucz (1999) web site for 
synthetic spectrum analysis of selected portions of the CH 
$A^{\rm 2}\Delta-X^{\rm 2}\Pi$ G-band and the CN 
$B^{\rm 2}\Sigma^+-X^{\rm 2}\Sigma^+$ violet bands.
With these linelists we derived a clear abundance enhancement of C 
in \oneonefive\ relative \onetwotwo\ (Table~\ref{tab4-abund},
Figure~\ref{fig5-diffabund}).
This agrees with the initial findings of GGGV82, who suggested
that [C/Fe]~$\simeq$~0 in \oneonefive\ while [C/Fe]~$\simeq$~--0.4
in \onetwotwo. 
Furthermore, although the [C/Fe] estimates of Kraft \etal\ 
(1982)\markcite{Ketal82} are nearly identical for these two stars, 
so are their [Fe/H] estimates.
If we translate their and our abundances to [C/H] values, we agree 
that [C/H]~=~--3.1 and --3.2 for \oneonefive\ and \onetwotwo\, respectively.
Finally, both Sneden (1973)\markcite{Sn73} and Norris \etal\ (1997a) 
also derived [C/H]~$\simeq$~--3.1 for \onetwotwo.
All the analyses for C are in agreement.

{\it The CNO-group: nitrogen.} 
The only detectable CN-violet absorption in these stars occurs at the
(0,0) vibrational bandhead at \wave{3883}.
From the synthetic/observed spectrum comparisons we estimate 
[N/Fe]~$\simeq$~+1.2 for \oneonefive\ and +1.1 for \onetwotwo,
or [N/H]~$\simeq$ --1.8 and --1.7, respectively.
The estimated abundance uncertainty in fitting the very weak CN bandhead is 
$\simeq\pm$0.25~dex.
The Kraft \etal\ (1982) nitrogen abundances are somewhat smaller,
[N/H]~$\simeq$ --2.3 and --2.2.
Sneden (1973) obtained a larger abundance for \onetwotwo,
[N/H]~$\simeq$ --1.5, while the Norris \etal\ (1997a) value
for this star is nearly identical to ours.
A renormalization probably needs to be applied to either the
Kraft \etal\ N abundances or to the other studies
(or maybe to both sets of N abundances).
Resolution of this question is beyond the scope of our work.
A substantial improvement would be the observation of the 
NH $A^{\rm 3}\Pi_i-X^{\rm 3}\Sigma^-$ bands near \wave{3360}.
But the relative N abundance between the two stars is
likely to remain quite small in future analyses.

{\it The CNO-group: oxygen.} 
For the oxygen abundance determination we used the [\nion{O}{i}]
line at \wave{6300.3}. 
This line and the much weaker [\nion{O}{i}] \wave{6363} line are the
usual oxygen abundance transitions of choice for red giants.
We derived [O/Fe]~=~+0.66 for \oneonefive, and [O/Fe]~=~+0.61
for \onetwotwo. 
The latter value is in excellent agreement with the early 
work on \onetwotwo\ by Lambert, Sneden, \& Ries (1974)\markcite{LSR74}:
[O/Fe] = +0.6. 

{\it Other light elements.} 
The ``pure'' $\alpha$-capture elements Mg, Si, and Ca
have nearly identical normal UMP-star overabundances in both stars.
But caution should be exercised in interpreting the Si abundance,
as it is based on only one strong \nion{Si}{i} line at \wave{4102.94}
Another potentially useful \nion{Si}{i} line at \wave{3905.53}
is severely blended by CH \wave{3905.67}; the higher excitation
\nion{Si}{i} lines in the yellow-red were undetectably weak on our spectra.
The Al abundance rests solely on \nion{Al}{i} \wave{3961.53}
resonance line, as its companion at \wave{3944} suffers CH blending
(Arpigny \& Magain 1983\markcite{AM83}).
The derived Al underabundances in both stars [Al/Fe]~= --0.36 for \oneonefive\
and -0.29 for \onetwotwo) are however in good 
agreement with other values for UMP stars (Cayrel 1996).
The Na abundance is based on the \nion{Na}{i} doublet at
$\lambda\lambda$8183.2,8194.8~\AA, which occurs amid a thicket of
telluric H$_{\rm 2}$O lines.
Our EWs for the \nion{Na}{i} lines may not be as reliable as most
other transitions.
But the present [Na/Fe]~= 0.00 for \oneonefive\ and --0.15 for \onetwotwo\
are in good agreement with those derived by Pilachowski \etal\
(1996): +0.04 and --0.05, respectively.
The Pilachowski \etal\ abundances were based on spectrum syntheses of the 
$\lambda\lambda$5682.6,5688.2~\AA\ doublet, and give no indication that
that our Na abundances from the longer wavelength lines are in error.

{\it Fe-peak Elements.}
The Fe-peak abundance pattern in our stars (Figure~\ref{fig4-metallicity}) 
follows the general distribution in other UMP stars, and \oneonefive\
generally cannot be distinguished from \onetwotwo.
A conspicuous exception to this statement is manganese, for which
we derive very different abundances in the two stars: 
$\Delta$~\eps{Mn}~=~--0.49, more than 4$\sigma$ smaller than
$\Delta$~\eps{Fe}. 
The Mn anomaly is easily seen from inspection of the data: whereas 
typical weak Fe-peak absorption features in the \oneonefive\ 
spectrum are about half as strong as those in the \onetwotwo\ 
spectrum, the weak \nion{Mn}{i} lines are a third as strong
(Table~\ref{tab2-EWs}).
The Mn result is not an artifact of neglect of hyperfine splitting in 
the analyses, as we performed spectrum syntheses 
with the full hfs substructure for the strong 
$\lambda\lambda$4030.7,4033.0,4034.4~\AA\ lines.  
The relative Mn abundance difference between these two stars
therefore seems to be real. 
A future check on this result ought to be the observation of \nion{Mn}{ii}
lines in the \wave{3400} spectral region. 
For copper there are extreme deficiencies in both stars 
([Cu/Fe]~$\simeq$~--0.7), but this is not a new result; all UMP
stars apparently have very depressed levels of copper (Sneden, Gratton,
\& Crocker (1991)\markcite{SGC91}.

Titanium (probably belonging to both $\alpha$-capture and Fe-peak
element groups) is overabundant, but much more so in \oneonefive;
the mean of \nion{Ti}{i} and \nion{Ti}{ii} relative abundances is
$\Delta$~\eps{Ti}~=~--0.06, which is about 3$\sigma$ higher
than $\Delta$~\eps{Fe}.
As with the case of Mn, the Ti anomaly can be readily forecast by 
inspection of the spectra of the two stars.
The Ti abundance difference is not easily explained, as it is based on 
large numbers of both neutral and ionized species transitions.
Departures from LTE, which are conceivable as such, probably cannot
account for this abundance difference, since we must demand that any 
proposed non-LTE corrections should affect neutral and ionized 
species abundances in significantly different ways in the two stars.

\subsection{The Neutron-Capture Elements}

The abundances of the neutron-capture elements (Z~$>$~30) were determined
by EWs or by synthetic spectra, as indicated in Table~\ref{tab2-EWs}. 
The linelists for these calculations were those originally developed 
for the abundance analysis of CS~22892-052 (Sneden \etal\ 1996).
Hyperfine structure was explicitly accounted for in the abundance
computations for Ba, La, and Eu; see Sneden \etal\
for further discussion and references to the laboratory data.
An example of a synthetic spectrum fit to the same features in both stars 
is shown in Figure~\ref{fig6-synspectra}.
It is clear that the derived abundance from the \nion{Eu}{ii} feature 
in \oneonefive\ is greatly enhanced relative that in \onetwotwo. 
As \oneonefive\ is more metal-poor than \onetwotwo, the contrast between
the europium abundance is very large between the two stars. 
This statement holds for all of the heavier (Z~$\geq$~56) elements
(Table~\ref{tab4-abund}, Figure~\ref{fig5-diffabund}).
On the other hand, the lighter neutron-capture elements 
(32~$\leq$~Z~$\leq$~40) are only weakly enhanced in \oneonefive.
These results were anticipated from the spectra presented in 
Figure~\ref{fig1-sample}, and are in qualitative agreement with the
earlier studies of GGGV82 and Gilroy \etal\ (1988).

With this work we have increased the total number of neutron-capture
element abundances in \oneonefive\ to 20.
The reader is reminded that the abundances for Ge, Os, and
Pt were obtained from comparing extant HST GHRS observations in 
three small UV spectral intervals to synthetic spectrum computations with
the newly derived models.
As expected, the new abundances do not differ significantly from those
of Sneden \etal\ (1998a)\markcite{SCBT98}, because the new model parameters
are not significantly different from the ones used in that study (the
new models are about 100~K cooler in \teff\ and about 0.15~dex lower
in log~g).
Sneden \etal\ urged caution in interpretation of their lead abundance,
which was derived from one weak \nion{Pb}{i} line (possibly corrupted by
a noise spike) at \wave{2833.05}.
Our new inspection of that line re-affirmed the earlier concerns,
and so we decided not to include Pb among the neutron-capture elements
in our present work.

\subsection{Thorium and Uranium}

The \nion{Th}{ii} \wave{4019.1} line is the only thorium feature that
has been employed for abundance determination in metal-poor stars, 
although recently Sneden \etal\ (1998a) have detected the weaker 
\wave{4086.5} line in the extremely r-process rich star CS~22892-052.
The \wave{4019} line is heavily blended, and several previous studies
(\eg, Morell \etal\ 1992\markcite{MKB92}, Fran\c{c}ois \etal\ 
1993\markcite{FSS93}, Norris \etal\ 1997b\markcite{NRB97b}) have 
carefully considered possible contaminant lines to the total feature.
A preliminary Th abundance in \oneonefive\ has been presented
by Cowan \etal\ (1999a) from a spectrum combined from our 2d-coud\'e
data and similar data from the Keck~I HIRES.
Here we discuss the analysis of just the 2d-coud\'e data.

We carried out synthetic spectrum computations for both stars,
varying several elemental and CH abundances in order to get the
best overall match to the observed feature.
In Figure~\ref{fig7-thorium} we show the results of these experiments.
In each panel, syntheses with different thorium abundances are displayed.
The solid line represents the case where Th is deleted from the analyses. 
To set the strengths of the $^{\rm 13}$CH lines, we first determined
carbon isotope ratios from synthetic spectra of the 
$A^{\rm 2}\Delta-X^{\rm 2}\Pi$ band lines near \wave{4230} and \wave{4370},
obtaining \carbiso~= 4.0~$\pm$1.5 for \onetwotwo\ and 6.0~$\pm$1.5
for \oneonefive\ (in agreement with earlier studies by Lambert \& 
Sneden 1977\markcite{LS77} and Sneden \etal\ 1986\markcite{SPV86}).
Then we empirically determined the CH abundance that would best match 
the observed $^{\rm 12}$CH $B^{\rm 2}\Sigma-X^{\rm 2}\Pi$ lines surrounding
\wave{4019} (for example, the $\lambda\lambda$4020.0,4020.2~\AA\ CH pair
shown in Figure~\ref{fig7-thorium}).
Finally, the strengths of the $^{\rm 13}$CH 
$\lambda\lambda$4018.97,4019.14~\AA\ lines were scaled down by the
\carbiso\ ratio.

For \onetwotwo\ the entire \wave{4019} absorption may be satisfied 
without any contribution from the \nion{Th}{ii} line 
(Figure~\ref{fig7-thorium}, bottom panel).
We can only suggest that \eps{Th}~$\lesssim$~--3.1 in this star.
The lack of Th detection is however consistent with the low abundances
of all other neutron-capture elements in \onetwotwo.
The \nion{Th}{ii} line is cleanly detected in \oneonefive; it dominates
the rest of the absorption near \wave{4019.1} (Figure~\ref{fig7-thorium},
top panel).
The best-fit synthetic spectrum yields \eps{Th}~= --2.23~$\pm$~0.05, 
where the quoted error is a measure of uncertainty in the fitting procedure.
This value is to be preferred to the preliminary values estimated
by Cowan \etal\ (1999a\markcite{Cetal99}) of \eps{Th}~= --2.10~$\pm$~0.1. 
Cowan \etal\ also unsuccessfully tried to detect the \wave{4086} 
\nion{Th}{ii} line.
We concur here with their assessment that from this feature, 
\eps{Th}~$\leq$~--2.2, not inconsistent with the detected
\wave{4019} line.

Cowan \etal\ (1999a) also employed their combined spectrum to search for
features of uranium in \oneonefive.
The most promising line of \nion{U}{ii} occurs at \wave{3859.58},
but in most cool stars it is hopelessly lost amid the strong CN
absorption in this spectral region.
But since CN is extremely weak in \oneonefive, a search for this
line is feasible.
However, Cowan \etal\ and now we fail to detect the line (see their
Figure~6), and the upper limit we derive from the 2d-coud\'e spectrum 
is \eps{U}~$<$~--2.6.
All other \nion{U}{ii} lines in the visible and near-UV spectral
lines should be undetectably weak in \oneonefive.

\section{Discussion}

Our abundance analyses have indicated (1) that the heavier neutron-capture 
elements are overabundant by a factor of 10 in \oneonefive\ compared 
to \onetwotwo.
(2) The iron-peak elements have similar abundances in \onetwotwo\ and 
\oneonefive, with the notable exceptions of Ti and Mn. 
(3) The carbon abundance is high in \oneonefive. 
We discuss these points in more detail in this section.

\subsection{Comments on the CNO Abundances}

Nearly all high luminosity (\MV~$\lesssim$~0) halo giant stars exhibit
very low \carbiso\ ratios and low total carbon abundances, 
indicating efficient operation of CN-cycle nucleosynthesis and 
convective envelope dredge-up along the giant branch.
The required envelope mixing is much more than that predicted for
the canonical first dredge-up in such low mass, low metallicity stars.
The extra mixing almost surely begins at the red giant stage in
which the outwardly advancing H-burning shell contacts the molecular
weight barrier left behind by the first-dredge up.
Recent theoretical studies of this phenomenon have been conducted by 
Charbonnel (1995)\markcite{Ch95}; new observational confirmation and 
references to prior work are given by Gratton \etal\ (1999)\markcite{GSCB99}.

Our stars match well the general trends of CNO abundances in low 
metallicity giants.  
But the carbon abundance in \oneonefive\ relative to \onetwotwo\ is large 
($\Delta$~\eps{C}~= +0.10, compared to $\Delta$~\eps{Fe}~= --0.24).
Moreover, \onetwotwo\ has a smaller isotopic ratio (\carbiso~$\simeq$~4) 
than does \oneonefive\ (\carbiso~$\simeq$~6), and 
it also has a smaller \eps{C}. 
Therefore one obvious possibility is that \onetwotwo, at a more 
advanced giant branch evolutionary state (lower \teff, lower \MV)
than \oneonefive, has had a chance to mix more CN-cycle products 
to its surface.
The C abundances with respect to iron are [C/Fe]~=~--0.11 
for \oneonefive, and [C/Fe]~=~--0.46 for \onetwotwo.
Mean C abundances in low metallicity giants average [C/Fe]~$\sim$~0.0, 
but individual cases vary widely: some stars have [C/Fe]~$\sim$~--0.5, and
others have extreme overabundances, [C/Fe]~$\gtrsim$~+1.0 
(\eg, McWilliam \etal\ 1995, Norris \etal\ 1997a).
The McWilliam \etal\ carbon abundances for stars with 
--2.5~$\geq$~[Fe/H]~$\geq$~--3.0 naturally divide into low-carbon
($<$[C/Fe]$>$~=~--0.48~$\pm$~0.08) and high-carbon 
($<$[C/Fe]$>$~=~+0.14~$\pm$~0.06) groups.
The low-carbon group of six stars has $<$log~g$>$~=~0.9~$\pm$~0.1,
and the high-carbon group of eight stars has $<$log~g$>$~=~2.0~$\pm$~0.2.
Thus \oneonefive\ and \onetwotwo\ seem to have [C/Fe] 
ratios consistent with their metallicities and gravities.

An alternate interpretation might link large carbon abundances with
large neutron-capture abundances in UMP giants, because [C/Fe]~=+1.0 
in the extremely $r$-process-rich star \cs22892.
In this view, the more mild enhancement of the $r$-process abundances
in \oneonefive\ has been accompanied by a much smaller carbon enhancement.
To test this idea, we plotted the [Eu/Fe] ratios for a sample of 28
of field halo giants (from Burris \etal\ 1999) against their [C/Fe] 
ratios (from Kraft \etal\ 1982).
The result was a scatter diagram.
Nor did a correlation emerge when newly derived [C/Fe] ratios for 27 of
these stars (Rossi \etal\ 1999)\markcite{RBS98} were correlated with the
[Eu/Fe] values.

At present, differences in convective envelope dredge-up efficiency 
seems the more likely explanation for the difference in carbon 
abundance of our two stars.
We would like to ascribe the slightly higher \carbiso\ ratio of 
\oneonefive\ to this phenomenon also, but the observational uncertainties
here are too large to support such a claim.
Determination of nitrogen abundances for the McWilliam \etal\ (1995)
sample could illuminate this question further.
However, the observed [(C+N)/Fe] abundance sums for these stars are much 
larger than those that would be expected if they began their lives with 
[C/Fe]~= [N/Fe]~= 0.0 (as is observed in metal-poor main sequence stars; 
see \eg\ Laird 1985\markcite{La85}, Carbon \etal\ 1987\markcite{CBKFS87}, 
Tomkin \& Lambert 1984\markcite{TL84}, Tomkin, Sneden, \& Lambert 
1986\markcite{TSL86}).
This cannot be resolved until nitrogen abundance scale questions
are satisfactorily understood.

The derived oxygen abundances, [O/Fe]~$\simeq$~+0.6, are among the largest
reported values for metal-poor giants from analyses of the 
[\nion{O}{i}] \wave{6300} line.
Previous surveys of such stars (Gratton \& Ortolani 1986\markcite{GO86}, 
Barbuy 1988\markcite{Ba88}, Gratton \etal\ 1999)
have found $<$[O/Fe]$>$~$\simeq$ +0.35 for 25 stars in the
metallicity range --1.0~$\geq$~[Fe/H]~$\geq$~--2.5.
Thus our larger [O/Fe] values may provide some support for the
recent suggestions (Israelian \etal\ 1998\markcite{IGR98},
Boesgaard \etal\ 1999\markcite{BKDV99}) that [O/Fe] increases with 
decreasing [Fe/H].
But the present study lacks appropriate control stars at higher metallicities;
caution must be observed as systematic observational/analytical errors 
can produce shifts between different studies.
This point should be pursued with uniform data sets of very high
S/N spectra in future studies.

\subsection{Abundances of the Iron-Peak Elements}

Recall from Figure~\ref{fig5-diffabund} that the relative abundances
$\Delta$~\eps{X} of nearly all Fe-peak elements are mostly 
consistent with the relative iron abundance metallicity difference 
$\Delta$~\eps{Fe} between the two stars.
Given this unity (based on eight elements), the relative manganese abundance 
abundance  $\Delta$~\eps{Mn}~$\simeq$~--0.5 is quite different.
Both stars are, however, within the range in [Mn/Fe] found for other 
stars of similar metallicity ({\it cf.} McWilliam \etal\ 1995).
Nakamura \etal\ (1999) have suggested that variations in the mass 
cut in Type~{\sc II} SNe explosions could lead to low values 
of certain elements such as [Mn/Fe] in low-metallicity stars. 
A deeper mass cut would expel larger amounts of complete Si-burning 
products (\ie\ Fe), but the same amount of incomplete Si-burning products 
(such as Mn), yielding net lower ratios of [Mn/Fe]. 
Such SNe with deeper mass cuts would also presumably eject more neutron-rich 
(\ie\ $r$-process) material consistent with the abundances in \oneonefive.
However, the Nakamura \etal\ results suggest that for a deep SN 
mass cut not just Mn, but also other products of incomplete Si-burning, 
such as Cr, should show the same trend of low values with respect to Fe, 
while the products of complete Si-burning, such as Co, should behave in the 
opposite manner.
Our Cr and Co abundances do show such effects, but they are nearly
identical in both stars; the Cr and Co abundances do not indicate a mass-cut
difference in SNe progenitors to \oneonefive\ and \onetwotwo.
Thus, while the depth of the mass cut might suggest a possible 
correlation between low Mn and high $r$-process element abundances
in \oneonefive, either some further quantitative exploration of the Nakamura 
\etal\ ideas or an alternative explanation for the low Mn abundance might 
be required.

The mild relative overabundance of Ti in \oneonefive, 
$\Delta$~\eps{Ti}~=~--0.06, also does not seem to be predicted
from inspection of the figures in Nakamura \etal.
The origin of this abundance difference may be due to differences in 
total (and/or core) masses of the probably relative small sample
of supernovae that once produced the heavy elements in the two stars. 
This suggestion is supported by recent early galactic nucleosynthesis
simulations by Karlsson \& Gustafsson (1999)\markcite{KG99}. 
Their models yield increasing scatter in [Ti/X], as well as in many other
abundance ratios, as overall Fe metallicity decreases from
[Fe/H]~=~--2.5 to --3.

\subsection{Abundances of the Neutron-Capture Elements}

The abundances of the elements Ba$\leftrightarrow$Dy in \oneonefive\
are significantly larger ($\sim$~10x) than in \onetwotwo,
even though the iron metallicity is very similar in these stars 
(Table~\ref{tab4-abund}, Figure~\ref{fig5-diffabund}).
This gross difference between the absolute levels of the heaviest 
neutron-capture elements in these two stars is one more clear example 
that the abundances of the heavier elements vary greatly with respect 
to Fe in UMP stars of the early galaxy (Gilroy \etal\ 1988; 
Burris \etal\ 1999). 
This presumably reflects the fact that the early Galaxy was not well mixed,
with some stars receiving larger absolute amounts of neutron-capture
material from individual nucleosynthetic events (almost certainly
Type~{\sc II} SNe).

Figure~\ref{fig8-rphd115} shows the newly-determined
neutron-capture element abundances for \oneonefive\ along with the 
HST abundances of Sneden \etal\ (1998a). 
Superimposed upon the abundance data is the scaled solar system 
$r$-process elemental abundance distribution.
The solar system $r$-process curve has been scaled vertically downward to
account for the lower metallicity of this star with respect to the Sun.
This elemental curve was obtained by summing over the isotopic
abundances, and these were based on deconvolving the total solar 
system abundances (Anders \& Grevesse 1989\markcite{AG89}) into 
{\it s}- and {\it r}-process abundance fractions, using the neutron 
capture cross sections of K{\"a}ppeler \etal\ (1989) and Wisshak, Voss, 
\& K{\"a}ppeler (1996)\markcite{WVK96}; see Sneden \etal\ (1996) 
and Burris \etal\ (1999) for details of this procedure.

The agreement between the heavier elements (Z~$\ge$~56) and the 
scaled solar system curve is very striking in Figure~\ref{fig8-rphd115}. 
While there has been increasing evidence that this pattern is 
true for elements in the range 56~$\leq$~Z~$\leq$~70, 
there have been few data for even heavier stable elements 
(76~$\leq$~Z~$\leq$~83, Os$\leftrightarrow$Bi, also know as
\third\ $r$-process peak).
The new observations of \oneonefive\ confirm the scaled solar system 
$r$-process pattern throughout the range 56~$\leq$~Z~$\leq$~78.
This in turn argues for a common site for $r$-process synthesis of these
elements.
Possible sites include supernovae and neutron-star binaries; see 
Wheeler, Cowan, \& Hillebrandt (1998)\markcite{WCH98} and 
Freiburghaus \etal\ (1999)\markcite{Fetal99} for an extended 
theoretical discussion.
But the observations will not easily support a wide range of
allowable $r$-process parameters: a narrow range of progenitor star
masses (\eg, Mathews, Bazan, \& Cowan 1992\markcite{MBC92}; 
Wheeler \etal\ 1998) or a narrow range of nuclear and astrophysical 
conditions in the $r$-process sites (\eg, Freiburghaus \etal\ 1999) 
are required.

The lighter neutron-capture elements Ge, Sr, Y and Zr shown in 
Figure~\ref{fig5-diffabund} present a puzzle, as their abundances are 
very similar in both stars.
This suggests that the abundances of these elements are dependent upon 
metallicity, unlike the heavier elements (see also McWilliam \etal\ 1995,
Sneden \etal\ 1998a, Burris \etal\ 1999, Cowan \etal\ 1999a).
Moreover, these elements in both stars are poorly matched by the
scaled solar system $r$-process curve in Figure~\ref{fig8-rphd115}.
Further observational evidence supporting this point comes from the 
identification of several elements in the 41~$\leq$~Z~$\leq$~48 range from 
new near-UV spectra of \cs22892\ (Cowan \etal\ 1999b)\markcite{CSIBBF99}.
The scaled solar-system $r$-process curve is also not a good fit to
the abundances of that more extensive set of elements with Z~$<$~56.
These mis-matches might demonstrate the existence of such a 
``second $r$-process'' that synthesizes these lighter $r$-process nuclei 
in a different environment than the heavier $r$-process nuclei
(see Wasserburg, Busso, \& Gallino 1996\markcite{WBG96}; 
Qian, Vogel, \& Wasserburg 1998\markcite{QVW98}). 
More extensive theoretical abundance predictions of these elements
(see \eg, Gallino \etal\ 1999\markcite{GBLTAV99}) are clearly needed here.

We have made quantitative comparisons of the abundances of elements 
with 56~$\leq$~Z~$\leq$~78 in \oneonefive\ and \cs22892\ to the 
solar system $r$-process abundance distribution.
Here, these elements will be collectively referred to as 
``heavy-r'' elements.
The weighted mean abundance offset of the heavy-r elements for 
\oneonefive\ is $<$[heavy-r/H]$>$~=~--2.06~$\pm$~0.03 
($\sigma$~=~0.10; 14 elements), and for \cs22892\ 
$<$[heavy-r/H]$>$~=~--1.44~$\pm$~0.02 ($\sigma$~=~0.06; 16 elements).
In these calculations, the weight employed for each element's 
abundance was its observed line-to-line scatter standard deviation. 
However, weighting by the number of lines used, or assigning equal weights to
each element produced little change in the mean abundances.
These means represent the r-process abundance levels of the heavy
stable elements in the two stars.
In the top and middle panels of Figure~\ref{fig9-residuals} we
show the scatter about the means for \oneonefive\ and CS~22892-052.
The larger scatter for \oneonefive\ seen in the top panel is evidence of the
weaker absorption lines of neutron-capture species in this star compared
to those of \cs22892, which increased the analysis difficulties for
some elements.
For example, the Tm abundance for \oneonefive\ comes a single very
weak and blended \nion{Tm}{ii} transition at \wave{3700.2}, while in 
\cs22892\ four \nion{Tm}{ii} lines were analyzed by Sneden \etal\ (1996).
But in both stars it is clear that there is a single-valued abundance
offset from the heavy-r solar system distribution.

The bottom panel of Figure~\ref{fig9-residuals} shows the comparison
of a combined abundance set (\oneonefive\ plus \cs22892) with the
heavy-r solar system values.
The combined stellar set was formed by first normalizing each 
star's heavy-r abundances by the mean abundance described above
(\eps{X}$_{\oneonefive}$~+~2.06, and \eps{X}$_{\cs22892}$~+~1.44).
Then we computed straight means of the normalized abundances for
those elements analyzed in both stars, or else adopted the normalized
abundances of one star when there was none available for the other
star.
Thus the abundance of terbium and hafnium in the combined set comes only from
\cs22892, while the platinum abundance comes only from \oneonefive.
Comparing this normalized, combined data set to the solar-system
$r$-process distribution yielded 
$<$[heavy-r/H]$>$~=~0.00~$\pm$~0.02 ($\sigma$~=~0.08; 17 elements).
No element abundance in the combined data deviates by more than 0.13~dex
from the solar system distribution.
Figure 9 clearly demonstrates that the abundances of the heavy 
stable elements in \oneonefive\ and \cs22892\
from Z~=~56--78 are consistent with the solar system r-process abundances.
While this has been suggested previously for \cs22892\
(Sneden \etal\ 1996), we now have extensive neutron-capture element
abundance data, including 
that in the 3$^{\rm rd}$ r-process peak, for two stars. 
These new data, 
particularly the normalized and combined abundances of \oneonefive\ and 
CS 22892--052 (as shown in Figure 9) yield solid agreement with 
solar values, making other explanations for this abundance pattern
(e.g., Goriely \& Arnould 1997\markcite{GA97}; Goriely \& Clerbaux 
1999\markcite{GC99}) less likely. 
(See also further discussion in Freiburghaus \etal\ 1999.)
As more r-process-rich UMP stars are discovered and subjected to detailed
analysis, the accidental errors should decrease even further,
leaving any remaining deviations to mark either true departures from
the solar abundances or residual abundance analysis problems.

\subsection{Chronometers and Ages}

The detection of radioactive thorium in \oneonefive\ 
(Figure~\ref{fig7-thorium}) provides an opportunity to determine 
the radioactive age of this star. 
Consider first the chronometric age based solely on 
the HD~115444 abundances. 
To minimize errors in abundance determinations we have employed the 
ratio of thorium, produced solely in the $r$-process,
to the stable abundance level of europium, an element overwhelmingly 
produced by the $r$-process (see Sneden \etal\ 1996, Burris \etal\ 1999). 
While Cowan \etal\ (1999a) found a preliminary value of 
\eps{Th}~= --2.1 $\pm$ 0.07 (based on an older model stellar atmosphere), 
our new synthesis indicates a value of \eps{Th}~= --2.23 $\pm$ 0.07, 
for thorium and an observed ratio of \eps{Th/Eu} = --0.60 in \oneonefive. 
This value is very similar to that found previously for \cs22892, of 
\eps{Th/Eu}~= --0.66 (Sneden \etal\ 1996).  

The observed ratio of Th/Eu in \oneonefive\ can be compared to the initial
predicted value at time of formation in an $r$-process environment.
While the site for the $r$-process is still not well-defined,
abundances of the heaviest stable neutron-capture elements detected 
in \oneonefive\ and other stars can be used to constrain the 
predictions of theoretical $r$-process calculations. 
These same calculations can then be employed to predict the 
zero-decay age abundances of the radioactive elements Th and U.
Recently Pfeiffer \etal\ (1997) and Cowan \etal\ (1999a),
used the most massive stable solar isotopic abundances
({\it i.e}, \iso{Pb}{206,207,208}\ and \iso{Bi}{209})
to constrain the theoretical calculations.
They conclude that the best fits to the observed solar data are obtained 
using the nuclear mass models FRDM-HFB (finite range droplet with 
Hartree-Fock-Bogoliubov), the ETFSI-Q (Extended Thomas Fermi with 
Strutinsky Integral and Quenching) and the ETFSI-Q, lsq (least square fit
applied to ETFSI-Q). 
These models predict initial values of Th/Eu of 0.496, 0.546 and 0.48, 
respectively (see Cowan \etal\ 1999a for details).
Comparing these initial values with the observed stellar ratio
yield values of 13.7, 15.7 and 13.1~Gyr, with an average age
for \oneonefive\ of 14.2 Gyr.

We can make additional chronometric age determinations
by also considering the case of \cs22892\
and the mean abundances, as discussed in \S 4.3 and shown in Figure 9.
Not only do both stars have stable heavy r-process abundance
distributions that are (within observational errors) identical with each
other and with the solar system r-process distribution, but
both stars show very similar and significantly subsolar abundances 
of thorium. 
This strongly implies that both stars are of similar age.
For \oneonefive, therefore, we can compute a ratio of Th to the mean
abundance curve determined by all of the observed stable elements.
Taking that value at Z~=~63 (the position of Eu) suggests that
the observed Eu value should be slightly higher by 0.05 in 
log $\epsilon$ (comparable to our observational uncertainty) 
yielding a slightly different ratio of log~(Th/Eu)$_{mean}$~=~--0.65. 
A similar exercise with the observed Eu abundance in \cs22892\
suggests a similar correction in the opposite direction.
Using the observed (Th and Eu) values from Sneden \etal\ (1996), we find a
value of log~(Th/Eu)$_{mean}$~=~--0.61, which agrees with the value for
in \oneonefive, again within the observational errors.
Finally, the average of observed, corrected values for these
two UMP stars ($<$log~(Th/Eu)$_{mean}>$~=~--0.63) yields the values
reported by Cowan \etal\ (1999a): 15.2 (FRDM+HFB), 17.1 (ETFSI-Q)
and 14.5~Gyr (ETFSI-Q, lsq.),
with a mean age of 15.6~Gyr for both stars.

With the detailed observations for two stars at this point,
our analysis suggests a reduced (with respect to only one star)
observational error in log~$\epsilon$ of $\pm$ 0.05. 
Applying that abundance error to the 
mean age determination suggests age uncertainties of $+{2.3} \atop {-2.4}$
Gyr, based solely upon observational errors.
Extensive error analysis and discussion (Cowan \etal\ 1999a) suggests that the
calculations best fitting the solar data have typical theoretical 
uncertainties of $\pm$ 3~Gyr. 
If we assume an average observational uncertainty of 2.4 Gyr,
then the total error (added in quadrature for uncorrelated errors)
is approximately $\pm$~4 Gyr, resulting in a final value of 
15.6 $\pm$ 4 Gyr for the age of these two UMP stars. 

These calculations demonstrate both the power and the sensitivity of 
this method, with small changes in observed values or theoretical 
predictions changing the ages by the order of several Gyrs.
While we have reduced the observational errors somewhat as a result
of having two stars to analyze,  it will  be
difficult to reduce further the total error without additional 
stellar observations of thorium and further theoretical calculations
employing the most reliable mass formulae for nuclei far from
stability.
It is worth noting, however, that this radioactive-decay 
technique does not suffer from uncertainties in chemical evolution,
since the time-scale for such evolution in the early Galaxy 
is much shorter than the radioactive decay half-life (14 Gyr) of Th and 
the age of the Galaxy.

Additional supporting age information could be provided by the uranium
chronometer in \oneonefive, but we were only able to set an upper limit
on the abundance of this element of 
\eps{U}~$\leq$~--2.6. 
Here is an age estimation based upon that upper limit. 
In contrast to the case for Th, in the case of U
we have to take into account the two different isotopes,
with very different half-lives, that make up this element.
Theoretical calculations based upon the ETFSI-Q model 
(see Cowan \etal\ 1999a; Pfeiffer \etal\ 1997) 
predict that the initial U elemental abundance is
dominated by the shorter half-life isotope, $^{235}$U, while
presumably the elemental abundance in the metal-poor stars is
mostly determined by $^{238}$U. 
Ignoring the initial $^{235}$U production entirely (this isotope decays 
in much less than a Hubble time), and only comparing the initial 
time-zero $^{238}$U and the observed upper limit in \oneonefive\ 
gives a lower limit on the age of this star of $>$ 11 Gyr.
This value is already within the estimated error for the Th age estimate. 
Since the actual U abundance may be much below our upper limit, the
implied lower limit on the stellar age (based upon the non-detection 
of U) is in reasonable accordance with the Th age determination.

The thorium age estimates determined here, both for \oneonefive\ by itself 
and by considering both UMP stars, are consistent with  
our earlier age estimates for \cs22892, the (so far) only other 
metal-poor Galactic halo star for which abundances of Th and an 
extensive set of other neutron-capture elements have been determined
(Cowan \etal\ 1997, Pfeiffer \etal\ 1997, 
Cowan \etal\ 1999a). Further, these new chronometric age
determinations  agree well with recent globular cluster
age estimates 
(see Pont \etal\ 1998\markcite{PMTV98})
and cosmological age determinations of 14.9 $\pm$ 1.5 Gyr 
(Perlmutter \etal\ 1999\markcite{Petal99}) and 14.2 $\pm$ 1.7 Gyr
(Riess \etal\ 1998)\markcite{Retal98}.
While values of the Hubble constant near 70 km sec$^{-1}$ Mpc$^{-1}$ 
and a flat universe
would imply cosmological age estimates much less than this, most  
recent work has suggested a relatively low density universe, which  
along with that Hubble constant have 
suggested a corresponding universal age of 13 Gyr (see e.g., Freedman 1999), 
still consistent with our estimated
stellar chronometric ages. 
We note, however, that we have found thorium in only two stars so far,
and additional detections will be needed to strengthen this
technique of radioactive dating.

\section{Conclusions}

We have performed a detailed abundance analysis of the UMP field giant
\oneonefive.
In addition to treating \oneonefive\ alone in a standard model atmosphere
analysis of equivalent widths and synthetic spectra, we have made a
line-by-line abundance comparison with the well-known UMP giant \onetwotwo.
The differential study yields very accurate relative atmosphere parameters
and abundances that illuminate a number of similarities and sharp
disagreements in the chemical compositions of these two stars.
Here we summarize the main results, which confirm and greatly extend 
the initial spectroscopic investigations by GGGV82 and Gilroy \etal\ (1988).

\begin{itemize}

\item \oneonefive's mean Fe-peak metallicity is approximately 0.24~dex
lower than that of \onetwotwo.
It is a slightly less evolved star, with $\Delta$~\teff~$\simeq$ +150~K,
$\Delta$~log~g~$\simeq$~+0.2, and $\Delta$~\MV~$\simeq$~+0.6.
The absolute metallicity of \oneonefive\ is [Fe/H]~=~--2.99,
 in good agreement with GGGV82, but this number is somewhat dependent
on the adopted temperature scale, choice of model atmosphere,
and assumptions made in constructing the models.

\item For elements with Z~$\leq$~30, \oneonefive\ has a fairly typical
halo star abundance distribution.
But comparison to \onetwotwo\ reveals several significant differences
in abundance patterns.
Most notable are the large relative overabundance of carbon and
underabundance of manganese in \oneonefive, and
the smaller but significant overabundance of titanium.

\item The lighter neutron-capture elements (32~$\leq$~Z~$\leq$~40) are only
mildly enhanced in \oneonefive; the mean relative overabundance is
$<\Delta\mbox{[Ge, Sr, Y, Zr/Fe]}>$~= +0.18 compared to \onetwotwo.

\item The heavier neutron-capture elements (Z~$\geq$~56) in \oneonefive\ are 
very enhanced: $<\Delta\mbox{[Ba, La, Nd, Eu, Er, Yb/Fe]}>$~= +0.95.
The abundance pattern among these elements is consistent with a scaled
solar-system r-process distribution.

\item Thorium is underabundant relative to the other heavy neutron-capture
elements, as it is in \cs22892.
A simple age estimate for these two stars is 15.6~$\pm$~4~Gyr.

\end{itemize}
\acknowledgments

We thank Roberto Gallino, Inese Ivans, Francesca Primas, 
Friedel Thielemann, Jim Truran, Craig Wheeler and referee Roger Cayrel
for helpful discussions and comments on this paper.
Model atmospheres used in this paper were kindly computed by Bengt Edvardsson.
JW is grateful for financial support from the international student exchange 
program between Uppsala University and the University of Texas.
This research was funded in part by NSF grants AST-9618364 to CS and 
AST-9618332 to JJC.
Additional support was provided through grants GO-05421, GO-05856,
and GO-06748 from the Space Science Telescope Institute, which is
operated by the Association of Universities for Research in Astronomy,
Inc., under NASA contract NAS5-26555.

\clearpage

\begin{table}
\dummytable\label{tab1-basic}
\end{table}

\begin{table}
\dummytable\label{tab2-EWs}
\end{table}

\begin{table}
\dummytable\label{tab3-errors}
\end{table}

\begin{table}
\dummytable\label{tab4-abund}
\end{table}

\clearpage

\begin{center}
{\bf Figure Captions}
\end{center}
 
\figcaption{Two small spectral regions in \oneonefive\ and \onetwotwo.
Both panels illustrate the generally stronger lines of \onetwotwo,
caused by its lower \teff\ and larger metallicity.
The top panel shows the much deeper lines of \nion{La}{ii} and
\nion{Eu}{ii}, hinting at very large overabundances of the
heavier neutron-capture elements in \oneonefive.
But the bottom panel shows that the \nion{Zr}{ii} and \nion{Sr}{ii} 
lines of \oneonefive\ are actually weaker than those of \onetwotwo,
indicating that the lighter neutron-capture elements have comparable
abundance ratios with respect to Fe in both stars.
\label{fig1-sample}}

\figcaption{Abundances of \nion{Fe}{i} lines in \oneonefive\ plotted
as functions of excitation potential (EP) and reduced width
(log~EW/$\lambda$).
These abundances were generated with the final adopted model for \oneonefive.
\label{fig2-felines}}

\figcaption{Relative abundances between the two stars of \nion{Fe}{i}, 
\nion{Fe}{ii}, \nion{Ti}{i}, \nion{Ti}{ii}, and \nion{Ca}{i} lines.
These abundances were generated with the final adopted models.
\label{fig3-deltalines}} 

\figcaption{Individual metallicities for \oneonefive\ and \onetwotwo. 
The upper panel shows the abundances for the light and Fe-peak
elements and the lightest neutron-capture elements, and
the bottom panel shows the abundances of the heavier neutron-capture
elements.
The solid horizontal line at [Fe/H]~=~--2.74 represents the metallicity of 
\onetwotwo, and the dotted horizontal line at [Fe/H]~=~--2.99 represents
the metallicity of \oneonefive.
Element symbols are plotted at the appropriate atomic number and
most convenient [M/H] position near a data point.
\label{fig4-metallicity}} 

\figcaption{Differential abundances for \oneonefive\ and \onetwotwo. 
The definition of $\Delta$~\eps{X} identical to that of Figure 3.
The horizontal dashed line is placed at $\Delta$~\eps{Fe}~=~--0.24,
and solid horizontal lines are drawn at $\Delta$~\eps{X}~=--0.12 and
--0.36, representing $\pm$2$\sigma$ excursions from 
$\Delta$~\eps{Fe}.
See the text for further discussion of these points.
\label{fig5-diffabund}} 

\figcaption{Synthetic and observed spectra of the \nion{Eu}{ii} 
\wave{4129.8} line and other nearby neutron-capture features.
The observed data points are represented by open circles.
For \oneonefive (top panel), synthetic spectra have been generated for
\eps{Eu}~= --2.39, --1.61, --1.39, and --0.99,
while for \onetwotwo (bottom panel, the synthetic spectra are for
\eps{Eu}~= --2.99, --2.64, --2.19, and --1.89.
\label{fig6-synspectra}}

\figcaption{Synthetic and observed spectra of the \nion{Th}{ii}
\wave{4019.1} line and associated contaminant features.
The observed data points are represented by open circles.
The main blending species transitions have been labeled.
For \oneonefive (top panel), synthetic spectra have been generated for 
\eps{Th}~=  --$\infty$, --2.48, --2.23, and --1.98.
while for \onetwotwo (bottom panel, the synthetic spectra are for 
\eps{Th}~= --$\infty$, --3.08, --2.58, and --2.08.
In each panel the synthetic spectrum drawn as a solid line represents
the predicted total absorption without the presence of the \nion{Th}{ii}
line, and the dotted lines represents varying amounts of additional thorium.
\label{fig7-thorium}}

\figcaption{Neutron-capture Abundances in \oneonefive (filled circles). 
HST abundances from Sneden \etal\ 1998 are indicated by filled diamonds,
and for those elements the filled circles represent our re-analysis
of the HST data.
The line labeled ``SS Abundances'' in the figure legend is the 
scaled solar system $r$-process-only abundance curve. 
See the text for further explanation of the match to the solar system 
distribution.
\label{fig8-rphd115}}

\figcaption{Differences between normalized observed abundances
of elements with Z~$\geq$~56 to $r$-process abundances of these
elements in the solar system.
See the text for explanations of the normalization constant
indicated for the abundances displayed in each panel, and of the 
combination method for the abundances in the bottom panel.
Derived abundances of stable elements in this atomic number 
range are indicated by filled circles, and those of the radioactively
unstable element thorium is indicated with five-pointed stars.
\label{fig9-residuals}}

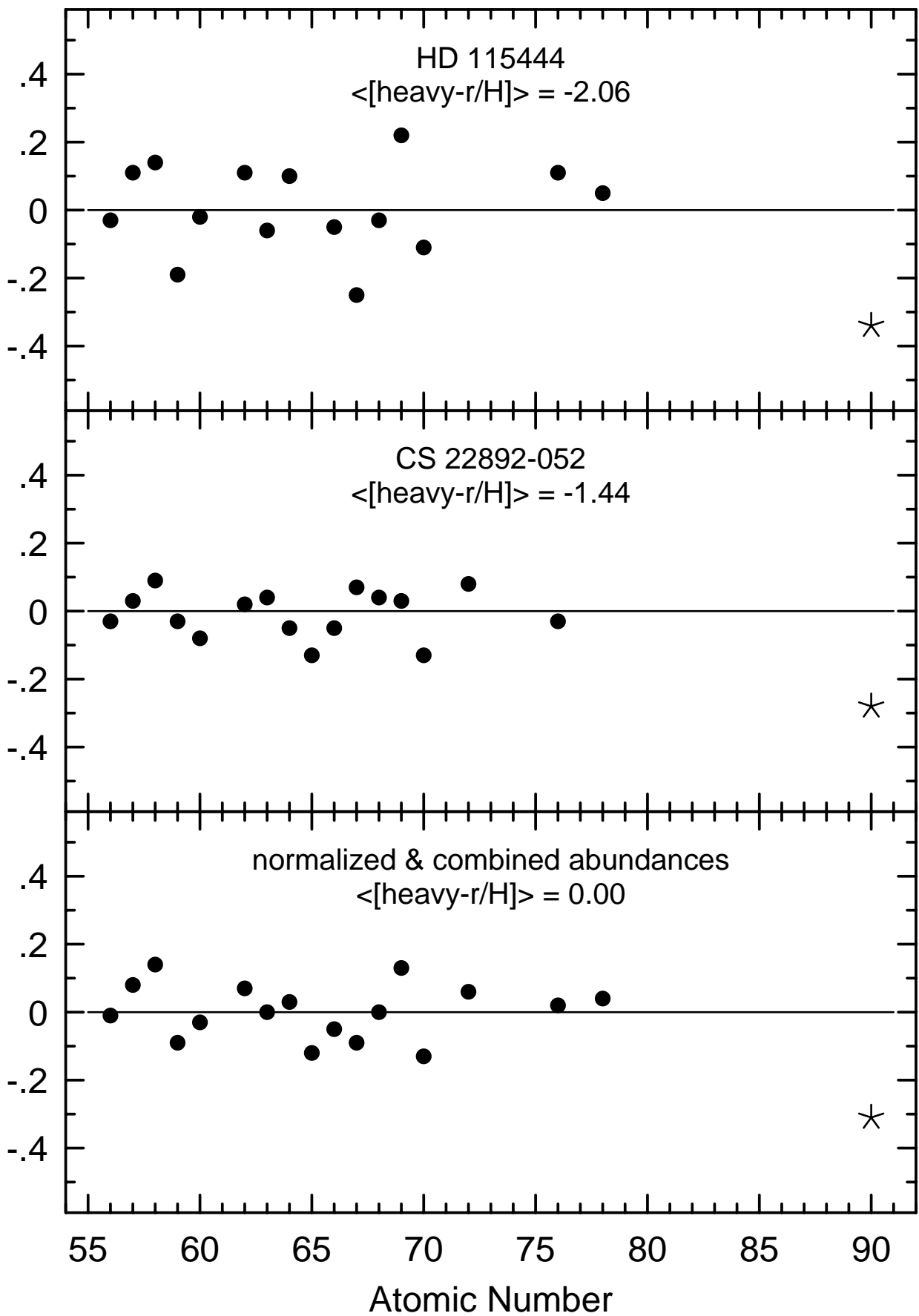
\end{document}